\documentclass[10pt]{article}
\usepackage{wrapfig,array}
\usepackage{epsfig,cancel,amsthm,amssymb, todonotes}
\usepackage{color,tikz}
\usetikzlibrary{decorations.markings,arrows, calc}
\usepackage{color}
\usepackage{graphicx,framed,verbatim,caption,amsbsy}
\usepackage[colorlinks=true, pdfstartview=FitV, linkcolor=darkblue, citecolor=darkblue, urlcolor=darkblue]{hyperref}
\definecolor{shadecolor}{rgb}{0.9, 0.9, 0.86}
\definecolor{darkgreen}{rgb}{0.2, 0.5,  0}
\definecolor{darkblue}{rgb}{0.1,0.1,0.45}

\def\&{\vspace{-5pt}&}

\def\Re{\mathrm {Re}\,}
\def\Im{\mathrm {Im}\,}

\def \eqref#1{(\ref{#1})}
\def \& {&\hspace{-10pt}}
\def\Ai{ {\mathrm {Ai}}}

\renewcommand{\d}{\mathrm d}
       
\newtheorem{theorem}{Theorem}[section]
\newtheorem{example}[theorem]{Example}
\newtheorem{exercise}[theorem]{Exercise}

\newtheorem{lemma}[theorem]{Lemma}
\newtheorem{remark}[theorem]{Remark}
\newtheorem{problem}[theorem]{Riemann-Hilbert Problem}

\newtheorem{proposition}[theorem]{Proposition} 
\newtheorem{corollary}[theorem]{Corollary} 
 
\newtheorem{definition}[theorem]{Definition}

\def\le{\left}
\def\ri{\right}
\def\ds{\displaystyle}

\def\bt{\begin{theorem}}
\def\et{\end{theorem}}
\def\bc{\begin{corollary}}
\def\ec{\end{corollary}}
\def\bx{\begin{example}}
\def\ex{\end{example}}
\def\bxr{\begin{exercise}\small}
\def\exr{\end{exercise}}
\def\bl{\begin{lemma}}
\def\el{\end{lemma}}
\def\bd{\begin{definition}}
\def\ed{\end{definition}}
\def\bp{\begin{proposition}}
\def\ep{\end{proposition}}

\def\br{\begin{remark}}
\def\er{\end{remark}}

\def\be{\begin{eqnarray}}
\def\ee{\end{eqnarray}}
\def \Tr {\mathrm{Tr}\,}
\def\ov {\overline}
\def\&{\hspace{-15pt}&}
\def\bea{\begin{eqnarray}}
\def\eea{\end{eqnarray}}
\def\beas{\begin{eqnarray*}}
\def\eeas{\end{eqnarray*}}

\def\R{{\mathbb R}}
\def\N{{\mathbb N}}

\def\wh{\widehat}

\def\Z{{\mathbb Z}}

\def\l{ \lambda }
\def\1{{\bf 1}}
\def\s{ {\sigma}} 
\def\z{\zeta}
\def\diag{{\mathrm{diag}}}

\def\QED {\hfill $\blacksquare$\par\vskip 3pt}

\makeatletter
\@addtoreset{equation}{section}
\makeatother

\begin{document}

\baselineskip 14pt plus 1pt minus 1pt

\begin{flushright}
\end{flushright}
\vspace{0.2cm}
\begin{center}
\begin{Large}

\textbf{
 Universality of the matrix Airy and Bessel functions at spectral edges of unitary ensembles} 
\end{Large}
\end{center}
\bigskip
\begin{center}
M. Bertola$^{\dagger\ddagger \clubsuit}$\footnote{Marco.Bertola@\{concordia.ca, sissa.it\}},  
M. Cafasso $^{\diamondsuit}$ \footnote{cafasso@math.univ-angers.fr}.
\\
\bigskip
\begin{minipage}{0.7\textwidth}
\begin{small}
\begin{enumerate}
\item [${\dagger}$] {\it  Department of Mathematics and
Statistics, Concordia University\\ 1455 de Maisonneuve W., Montr\'eal, Qu\'ebec,
Canada H3G 1M8} 
\item[${\ddagger}$] {\it SISSA/ISAS, via Bonomea 265, Trieste, Italy }
\item[${\clubsuit}$] {\it Centre de recherches math\'ematiques,
Universit\'e de Montr\'eal\\ C.~P.~6128, succ. centre ville, Montr\'eal,
Qu\'ebec, Canada H3C 3J7} 
\item [${\diamondsuit}$] {\it LAREMA, Universit\'e d'Angers\\ 2 Boulevard Lavoisier, 49045 Angers, France.}
\end{enumerate}
\end{small}
\end{minipage}
\vspace{0.5cm}
\end{center}
\bigskip
\begin{center}
\begin{abstract}
\noindent This paper deals with products and ratios of average characteristic polynomials for unitary ensembles. We prove universality at the soft edge of the limiting eigenvalues' density, and write the universal limit in function of the Kontsevich matrix model (``matrix Airy function'', as originally named by Kontsevich).\\
\noindent For the case of the hard edge, universality is already known. We show that also in this case the universal limit can be expressed as a  matrix integral (``matrix Bessel function'') known in the literature as generalized Kontsevich matrix model.
\end{abstract}
\end{center}

\section{Introduction}

One of the most interesting features of random matrices is universality. Universality was originally conjectured by Wigner and Dyson in the sixties, on the base of some considerations originating from statistical physics (see \cite{MehtaBook} and references therein). Roughly speaking, it predicts that, given a large size random matrix, the microscopical statistical behavior of its eigenvalues depends just on the class of symmetry of the matrix (Hermitian, symplectic, unitary...) and not on the particular features of the model in analysis.\\ 
\\
 
 In this paper we are interested in universality for the products and ratios of average characteristic polynomials of random matrices. Starting from the late nineties, these correlation functions had been thoroughly studied both from physicists and mathematicians. In physics, applications were found to the study of quantum chromodynamics (see \cite{OxRMBook}, chapters 19 and 32 and references therein) and quantum chaos (see for instance \cite{AndreevSimons}). In mathematics, average characteristic polynomials (for circular unitary ensemble) are closely related to the moments of the Riemann zeta function on the critical line, as shown in the seminal paper \cite{KeatingSnaith} (see also \cite{OxRMBook} and references therein).\\

Here we will consider unitary invariant ensembles on the space of Hermitian and positive semidefinite Hermitian matrices. More specifically, we analyze the following correlation functions 
\footnote{Actually our considerations extend to more general cases, but we decided to give details just for these ones in order to reduce technicalities, see also the Remark \ref{generalcase}}:
\begin{itemize}
	\item \underline{Soft Edge Case:} 
	\be\label{IntroSoft}
		\le\langle\prod_{j=1}^{2S} {\rm e}^{-\frac n 2  V(\xi_j)}  \det (\xi_j-M) \ri\rangle_{\mathcal H_n},
	\ee
	here the average is taken with respect to the measure $\d \mu(M) = {\rm e}^{-n\Tr V(M)} dM$ on the space of $n \times n$ Hermitian matrices.
	\item \underline{Hard Edge Case:}
	\be\label{IntroHard}
	 	\le\langle\prod_{j=1}^{2S}  \xi_j^\frac \nu 2 {\rm e}^{-\frac n 2 V(\xi_j)}  \det (\xi_j-M)^{\pm 1} \ri\rangle_{\mathcal H_n^+},
	\ee
	here the average is taken with respect to the measure $\d \mu(M) = M^{\nu}{\rm e}^{-n\Tr V(M)} dM$ on the space of $n\times n$ semi--positive definite Hermitian matrices, and $\nu > -1$.
\end{itemize}

In both cases, $V$ is assumed to be a ``regular'' potential, in the sense of \cite{DKMVZ} (see also below).
These correlation functions have been studied, for finite $n$, in several articles \cite{BrezinHikami00, FyodorovStrahov03, BaikDeiftStrahov03}, where it was shown that they can be expressed as finite--size determinants involving the orthogonal polynomials associated to the given matrix model (together with their Cauchy transforms). Those formulas are particularly suited to study the large $n$ limit; the first results, in the Gaussian case and both in the bulk and on the edges (soft and hard) goes back to \cite{BrezinHikami00bis}. In \cite{FyodorovStrahov03bis,FyodorovStrahovtris}, Fyodorov and Strahov proved the universality in the bulk and Vanlessen obtained analogous results for the hard edge \cite{VanlessenUniverBessel}. Some not completely rigorous results for the soft edge are contained in \cite{AkemannFyodorov}.\\ 

The present paper sets two  main goals: on the one hand, we fill the apparent gaps on the analysis of the soft edge. On the other hand, both for the case of the soft and hard edge, we will establish a sort of ``universal duality formul\ae'' stating that the universal large $n$ limit of \eqref{IntroSoft} and \eqref{IntroHard}, up to an explicitly known function, is equal to a matrix model with external potential encoding the position of the points $\xi_j$ and whose size is equal to $2S$. For the case of the soft edge, this matrix model is nothing but the Kontsevich one (i.e. the matrix Airy function \cite{Kontsevich}). For the hard edge, the model belongs to the family of the so--called generalized Kontsevich matrix models (known also as generalized Kazakov-Migdal-Kontsevich models, \cite{KMMM}\footnote{For special values of the parameters, this model is equivalent to the Brezin-Gross-Witten model \cite{MMS}.}).\\  
Our main results are formulated in the following two theorems, for the case of the soft and hard edge respectively. In the statement of the first one we use the equilibrium measure associated to a regular potential; the definition is given below, equations \eqref{ineqs0} \eqref{ineqs}.
\begin{shaded}
\bt
\label{thmsoft}
Let $V$ be a real analytic regular potential and $a$ a right endpoint of the support of the associated equilibrium measure.
Let $\xi_1\dots, \xi_{2S}$ be points of the form $\xi_j = a + C^{-1} n^{-\frac 2 3}  y_j^2$, where $C$ is an appropriate constant (see \eqref{zoomconst}) and $\Re y_j>0$.
Then
\be
\lim_{n\to\infty} \frac{C^{S^2} n^{\frac {2S^2}{3}}}{\ds\prod_{\ell=n}^{n+S-1} h_\ell} \le\langle\prod_{j=1}^{2S} {\rm e}^{-\frac n 2  V(\xi_j)}  \det (\xi_j-M) \ri\rangle_{\mathcal H_n}
=  \frac{\ds\prod_{j<k} (y_j+ y_k)}{  2^{2S} \pi^{S}\ds\prod_{j = 1}^{2S} \sqrt{y_j}} {\rm e}^{-\frac 23 \Tr Y^3} Z_{2S}^{Kont}(Y),
\label{duality1intro}
\ee
where $ Y = {\rm diag}(y_1,\dots, y_{2S})$ and $Z_{2S}^{Kont}$ is the matrix Airy function (or Kontsevich' integral) \cite{Kontsevich}
\be
Z_{2S}^{Kont}(Y) =  
\frac{ \ds \int_{\mathcal H_{2S}} \d  H {\rm e}^{ \Tr \le(i\frac {H^3}3 - {Y} H^2 \ri)}}{ \int_{\mathcal H_{2S}} \d H {\rm e}^{- \Tr \le( {Y} H^2\ri)}}.
\ee
\et
\end{shaded}
Here and below we denoted with $h_\ell$ the squared norm of the degree $\ell$ monic polynomial of the family orthogonal with respect to the measure induced by $V$. \\

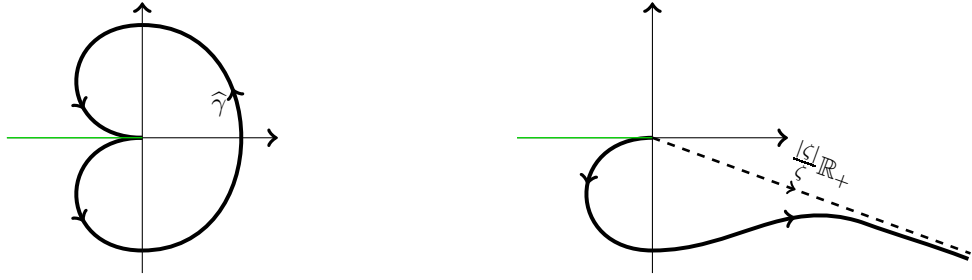
\begin{figure}[t]
\hfill
\begin{tikzpicture}[scale=1.5]
\draw [  postaction={decorate,decoration={markings,mark=at position 1 with {\arrow[line width=1.5pt]{>}}}}] (-1.2,0) to (1.2,0);
\draw [  postaction={decorate,decoration={markings,mark=at position 1 with {\arrow[line width=1.5pt]{>}}}}] (0,-1.2) to (0,1.2);
\draw [ line width = 1.5pt, postaction={decorate,decoration={markings,mark=at position 0.65 with {\arrow[line width=1.5pt]{>}}}}]
(0,0) to [out=180, in=180, looseness=2] (0,-1);
\draw [ line width = 1.5pt, postaction={decorate,decoration={markings,mark=at position 0.65 with {\arrow[line width=1.5pt]{>}}}}]
(0,-1) to  [out=0, in=0, looseness=1.5] node[pos=0.6, left]{$\wh \gamma$}(0,1);
\draw [ line width = 1.5pt, postaction={decorate,decoration={markings,mark=at position 0.65 with {\arrow[line width=1.5pt]{>}}}}]
(0,1) to [out=180, in=180, looseness=2] (0,0);
\draw [ green] (-1.2,0) to (0,0);
\end{tikzpicture}
\hfill
\begin{tikzpicture}[scale=1.5]
\draw [  postaction={decorate,decoration={markings,mark=at position 1 with {\arrow[line width=1.5pt]{>}}}}] (-1.2,0) to (1.2,0);
\draw [  postaction={decorate,decoration={markings,mark=at position 1 with {\arrow[line width=1.5pt]{>}}}}] (0,-1.2) to (0,1.2);
\draw [ line width = 1.5pt, postaction={decorate,decoration={markings,mark=at position 0.45 with {\arrow[line width=1.5pt]{>}}}}]
(0,0) to [out=180, in=180, looseness=2] (0,-1);
\draw [ line width = 1.5pt, postaction={decorate,decoration={markings,mark=at position 0.45 with {\arrow[line width=1.5pt]{>}}}}]
(0,-1) to [out=-0, in=162, looseness=1] (-22:2) 
to [out=-20, in=-201] (-21:3);
\draw [ green] (-1.2,0) to (0,0);
\draw [ dashed, line width = 1pt, postaction={decorate,decoration={markings,mark=at position 0.45 with {\arrow[line width=1pt]{>}}}}]
(0,0) to node[pos=0.5, above, sloped]{$\frac {|\z|}{\z}\R_+$}(-20:3);
\end{tikzpicture}
\hfill
\caption{The contour of integration for $f_\nu$ (left) and $g_\nu$ (right). The green color on $\R_-$ signifies the branch-cut of the integrand. As $\arg \z$ increases, the direction of integration for $g_\nu$ is deformed. The analytic continuation of $g_\nu$ requires to move the branch-cut of the integrand as well.}
\label{figBessel}
\end{figure}

The limit in the hard edge case involves integrals on slightly more complicated spaces of matrices: below we denote with $CUE_{\wh \gamma}$ the space of normal matrices of the form $U T U^\dagger$, with $U$ unitary, $T := {\rm diag}(t_1,\dots, t_{2S})$ and $t_j$ ranging on the contour $\wh \gamma$ defined below (see Figure \ref{figBessel}). If $\nu \in \N$ then we can also take $\wh \gamma$ as the unit circle and have the bona fide circular unitary ensemble (CUE). In this particular case, this matrix model corresponds to the one discussed in many different papers (see \cite{KMMM} and references above) as generalized Kontsevich matrix model. Here we prefer to call it \emph{matrix Bessel function}, in analogy with the soft hedge case.
\begin{shaded}
\bt
\label{thmhard}
Let $V$ be a real analytic regular potential and let $\xi_1\dots, \xi_{2S}$ be points of the form $\xi_j =  C^{-1} n^{-2 }  y_j$, with $C$ as in \eqref{zoomconst}.
Then
\be
\lim_{n\to\infty} 
&\& 
\frac{C^{S^2} n^{2S^2}}{ {\small \ds\prod_{\ell = n}^{n+S-1} \hspace{-2pt}h_\ell}}
 \le\langle\ds\prod_{j=1}^{2S}  \xi_j^\frac \nu 2 {\rm e}^{-\frac n 2 V(\xi_j)}  \det (\xi_j-M) \ri\rangle_{\mathcal H_n^{+}}
=\nonumber \\
&\&=
 \frac{ \det ({Y})^\frac \nu  2(2\pi)^S }{ \pi^{S(2S-1)}}
\int_{CUE_{2S,\wh \gamma}} \hspace{-22pt}(\det H)^{\nu-1}
{\rm e}^{ \Tr\le(- Y H + H^{-1}\ri)} \frac{\d H}{(2i\pi)^{2S}}\label{duality2intro}
\ee

\noindent where $Y = \diag(y_1,\ldots, y_{2S})$. Similarly, setting $\xi_j = -Cn^{-2} y_j$ with $y_j\not\in \R_-$, we obtain 
\be
\lim_{n\to\infty} 
\le(\prod_{\ell = n-S}^{n-1} \hspace{-8pt}h_\ell\ri)&\& 
C^{S^2} n^{2S^2}
 \le\langle\prod_{j=1}^{2S}  \frac{({\rm e}^{i\pi}\xi_j)^{-\frac \nu 2} {\rm e}^{\frac n 2 V(\xi_j)}}{  \det (\xi_j-M)} \ri\rangle_{\mathcal H_n^{+}}
=\nonumber \\
&\&=
 \frac{ \det ({Y})^\frac \nu  2(2\pi)^S }{ \pi^{S(2S-1)}}
\int_{\mathcal H^+_{2S}} \hspace{-12pt}(\det H)^{\nu-1}
{\rm e}^{ \Tr\le(-Y H - H^{-1}\ri)} {\d H}.\label{duality2IIIintro}
\ee
\et
\end{shaded}
\br
The integral in \eqref{duality2IIIintro} converges only for $\Re y_j>0$; if some of the $y$'s are taken in the right plane, we understand it as analytic continuation. 
\er

The name ``universal duality formul\ae'' is justified by the relation of our formul\ae\, with the well known duality, discovered by Brezin and Hikami \cite{BrezinHikami00}, reading
\be\label{dualityBH}
	\frac{1}{Z_n} \int_{\mathcal H_n} dM \prod_{i = 1}^k \det(\lambda_i - M){\rm e}^{-\frac{n}2 \Tr M^2 + N \Tr MA} = \frac{1}{Z_k} \int_{\mathcal H_k} dB \prod_{j = 1}^n \det(a_j - iB){\rm e}^{-\frac{n}2 \Tr(B - i\Lambda)^2}. 
\ee 
This formula exchanges $k$-points correlation functions on $(n\times n)$ matrices with $n$-points correlation functions on $(k \times k)$ matrices: note that here both the models are Gaussian with external potentials $A := \diag(a_1,\ldots,a_n)$ and $\Lambda := \diag(\lambda_1,\ldots,\lambda_k)$. As explained by the authors, taking as $A$ a multiple of the identity and computing the large $n$ limit of the equation above, one can prove that the large $n$ limit (on the edge) of $k$-points correlation functions for the Gaussian Unitary Ensemble (GUE) are proportional to the $(k\times k)$ Kontsevich matrix model. In other words, using \eqref{dualityBH} one can prove the equivalence between the Kontsevich matrix model and the ``edge of the spectrum'' matrix model \cite{OkounkovPandharipandeEdge} used by Okounkov and Pandharipande to prove the Witten--Kontsevich theorem\footnote{Very interestingly, one can generalize the formula \eqref{dualityBH} to beta models and, studying the asymptotics, express the $k$-points correlation functions of the Gaussian $\beta$ model in function of a $\beta$--generalization of the Kontsevich matrix model, see \cite{Desrosiers, DesrosiersLiu,DesrosiersLiubis}.}. For the Bessel case, a similar procedure had been used in \cite{BrezinHikamiBessel} to go from duality to the matrix Bessel function.\\

Unfortunately, the duality formula \eqref{dualityBH} exists just for the Gaussian case. However, our Theorems \ref{thmsoft}, \ref{thmhard} can be interpreted as stating that the duality gets restored, for arbitrary potentials, in the large $n$ limit (both in the case of the soft and hard edge) exchanging the large $n$ limit of $k$-points correlation functions (for models without external potentials) with partition functions for $(k\times k)$ matrix models (with external potential). From the point of view of universality, it is well known that the limit of the (rescaled) gap probability for Hermitian matrices, under the regimes described in Theorems \ref{thmsoft} and \ref{thmhard}, is given respectively by the Fredholm determinants of the Airy and the Bessel kernels. Here we prove that, analogously, the correlation functions of the same models, under the same regimes, are given respectively by the matrix Airy and Bessel functions. We plan to study other classes of universality in subsequent works.

\section{Finite $n$ correlators and their limit via Riemann--Hilbert problem}

It is a well established result that the averages of characteristic polynomials (for a given matrix model) can be expressed in terms of the solutions of a certain Riemann--Hilbert problem canonically associated to the corresponding orthogonal polynomials. Let us recall these results and fix the notations. In the following we denote with $\mu(z)$ a measure on $J = \mathbb R$ or $J = \mathbb R_+$ respectively of the form $\mu(x)\d x = {\rm e}^{-V(x)}\d x$ or $\mu(x) = x^{\nu}{\rm e}^{-V(x)}\d x$, with $V$ a regular potential and $\nu>-1$. As usual $\sigma_3$ denotes the Pauli matrix $\sigma_3 := \diag(1,-1)$.
\begin{problem}\label{RHProbOP}
Find a sectionally-analytic matrix--valued function $Y_n(z)$ on $\mathbb C/J$ satisfying the following jump and asymptotic conditions 
\bea\label{RHPOP}
Y_{n}(x)_+ &=& Y_n(x)_- \le[\begin{array}{cc}
1  &  \mu(x) \\
0 & 1
\end{array}\ri],\;
x\in  J,
\\
Y_n(z) &=& \le(\1 + \mathcal O(z^{-1})\ri) z ^{n\s_3}\ ,\ \ |z|\to \infty.
\eea
In the case  $J = \R_+$ we require the additional boundary behavior at $x = 0:$ 
\be
Y_n(z) =\le\{\begin{array}{cc}
  \mathcal O(|z|^{\min\{\nu,0\}}) & \nu \neq 0\\
  \mathcal O(\ln |z|) & \nu =0.
  \end{array}\ri.
\ee
\end{problem}
The solution is famously written in terms of orthogonal polynomials and their Cauchy transform as follows:
\bt
\label{OPRHPthm}{\cite{FIK0}}
The problem \ref{RHPOP} admits a unique solution of the form 
\be
Y_n(z) = \le[
\begin{array}{cc}
p_n(z) &\ds \frac 1{2i\pi } \int_J\frac  {p_n(x) \mu(x)\d x}{x-z}\\[15pt]
\ds\frac {-2i\pi}{h_{n-1}} p_{n-1}(z) &\ds \frac {-1}{h_{n-1} } \int_J\frac{p_{n-1}(x)\mu(x) \d x}{x-z}
\end{array}
\ri]
\ee
where $p_n$ are the  {\em monic} orthogonal polynomials for the measure $\ds\mu(x)\d x$ supported on $J$ and $h_n := \ds\int_J p_n^2(x)\mu(x)\d x$.
\et
Note in particular that the usual Christoffel--Darboux kernel is related to the above by the formula
\be\label{CDkernel}
K_n(x,y) = \frac {(Y_n^{-1}(x) Y_n(y))_{21}}{(-2i\pi) (x-y)}.
\ee
Averages of products and rations of characteristic polynomials can be expressed in terms of the solution of \ref{RHPOP} (i.e. in terms of the orthogonal polynomials and Cauchy transforms thereof) as in the next proposition. Note that, since formulas are valid for both cases, we do not specify if the average is taken on $\mathcal H_n$ or $\mathcal H_n^+$ but still we put the index $n$ to denote the size. As customary, we denote with $\Delta(\vec x)$ the Vandermonde determinant of the variables $\vec x = (x_1,\ldots,x_S)$. 
\bp[ \cite{FyodorovStrahov03}]
\label{finitecorr}
Let $b_i,c_j$ be arbitrary complex numbers, then
\bea
{\bf [I]}&\&\ \ \le\langle  \prod_{j=1}^S  {\det (b_j-M)}{\det(c_j-M)} \ri \rangle_n 
= 
 \frac {\ds\le(\prod_{\ell=n}^{n+S-1} h_\ell\ri) }{ (-2i\pi)^S \Delta(\vec c) \Delta(\vec b)} 
\det \le[
\frac {\big(Y^{-1}_{n+S}  (c_\ell ) Y_{n+S}(b_j)\big)_{21}}{c_\ell-b_j} 
\ri]_{1\leq \ell, j\leq S}\\
&\& =  \frac {\ds\le(\prod_{\ell=n}^{n+S-1} h_\ell\ri) }{\Delta(\vec c) \Delta(\vec b)} 
\det \Big[ K_{n + S}(c_\ell,b_j)
\Big]_{1\leq \ell, j\leq S}.\nonumber
\\[20pt]
{\bf [II]}&\&\ \ \le\langle  \prod_{j=1}^S \frac {\det (b_j-M)}{\det(c_j-M)} \ri \rangle_n = \frac{\prod_{j,\ell} (c_\ell- b_j)}{\Delta(\vec c)\Delta(\vec b)}
\det \le[
\frac {\big(Y^{-1}_{n}  (c_\ell ) Y_n(b_j)\big)_{11}}{c_\ell-b_j} 
\ri]_{1\leq \ell, j\leq S}
\\[10pt]
{\bf [III]}&\& \ \ \le\langle  \prod_{j=1}^S  \frac 1{\det (b_j-M) \det(c_j-M)} \ri \rangle_n =  
\frac {(-2i\pi)^S \ds \le(\prod_{\ell = n-S}^{n-1} h_\ell^{-1}\ri)}{\Delta (\vec c)\Delta (\vec b)}
\det \le[
\frac {\big(Y^{-1}_{n-S}  (c_\ell ) Y_{n-S}(b_j)\big)_{12}}{c_\ell-b_j} 
\ri]_{1\leq \ell, j\leq S}
\eea
\ep
\br
In \cite{FyodorovStrahov03} the formula for the third case is different (cf. (4.24) in loc. cit.) and contains a symmetrization operation. However the authors did not realize that the addenda are invariant under the symmetrization operation (this follows from Lemma \ref{LemmaDet} below), which yields the formula we gave above.
\er
\br All these formul\ae\ fall within the general framework of \cite{BertolaCafasso5}: indeed they are ratio of Hankel determinants of moments  of the measures $\mu(z)\d z$ and  $R(z) \mu(z)\d z$, with $R(z)$ an appropriate {\em rational} function (e.g. $R(z)  = \prod_{j=1}^{S} \frac {b_j-z}{c_j-z}$ in case $[II]$). 
This modification of the density $\mu$ corresponds to a conjugation of the jump matrix in the Riemann--Hilbert problem \ref{RHPOP} by a diagonal rational matrix, and hence the ratio of Hankel determinants is expressed as a determinant of the so--called {\em characteristic matrix}, which takes on the particular form exemplified above in these cases.\\ 
The different formul\ae\ of Uvarov-type in  \cite{BaikDeiftStrahov03} are also derived within the same framework by choosing different (equivalent)  ways of performing the conjugation of the jump matrix. 
\er

\subsection{Large $n$ asymptotics  near the edge of the spectrum}
\label{largen}
The aim of this subsection is to recall some of the results concerning the large $n$ asymptotics of the Riemann--Hilbert problem \ref{RHProbOP} and use them to compute the large $n$ limit of the formulas in Proposition \ref{finitecorr} . Accordingly, we will consider (for the case of the soft and hard hedge respectively) measures of the following forms:
\be
\mu(z) = \mu_n(z) := {\rm e}^{-n V(z)}\ ,\ \ z\in \R; \qquad
\mu(z) = \mu_n(z) := z^{\nu} {\rm e}^{-n V(z)}\ ,\ \ z\in \R_+, \ \ \ \nu>-1.
\ee
where $V$ is a real--analytic function of appropriate growth as $z\to\pm \infty$; 
common sufficient conditions are, respectively in the two cases: 
\be
\liminf_{|x|\to \infty} \frac {V(x)}{\ln |x|} = + \infty, \qquad
\hbox  {  and  }\qquad
\liminf_{x\to +\infty} \frac {V(x)}{\ln |x|} = + \infty, \ \ \inf_{x\in\R_+} V(x)>+\infty.
\label{grtg}
\ee
The results of \cite{DKMVZ} about the potential-theoretical properties of the potential $V$  can be summarized as follows:
\begin{enumerate}
\item There are, uniquely determined, a  real constant $\ell$ ({\em modified Robin constant}) and a positive measure (the {\em equilibrium measure}) with density $\rho$ and support consisting of a finite union of intervals such that the function ({\em effective potential}) 
\be\label{ineqs0}
\phi(x):= V(x) + \ell + \int_J \ln \frac 1{|x-t|} \rho(t) \d t
\ee
satisfies the inequalities 
\be
\phi(x) \equiv 0,\ \ \ \ x\in Supp (\rho)\ ,\ \ \ \phi(x)\geq 0\ , \ \ \ x\in  \le\{
\begin{array}{c}
\R \setminus Supp (\rho)\\
\R_+ \setminus Supp (\rho). 
\end{array}\ri.
\label{ineqs}
\ee
\item if $V$ is real analytic in an open  strip  around $\R$ ($\R_+$, respectively) and has the growth behaviours \eqref{grtg} then $\rho$ is analytic in the interior of the support and has the form 
\be
\rho(x) = M(x) \sqrt{|Q(x)|}\ ,\ \ \ \ Q(z):= \prod_{j=1}^{2s} (z-a_j)
\qquad a_1<a_2\dots < a_{2s},\\
\rho(x) = M(x) \sqrt{\frac{|Q(x)|}{|x|} }\ ,\ \ \ \ Q(z):= \prod_{j=2}^{2s} (z-a_j)\qquad 
0<a_2<\dots < a_{2s}
\ee
where the support of $\rho$ is  $[a_1,a_2] \cup [a_3,a_4] \cup \dots \cup [a_{2s-1}, a_{2s}]$ for the soft edge case and $[0,a_2]\cup \dots \cup [a_3,a_4] \cup \dots \cup [a_{2s-1}, a_{2s}]$ for the hard edge case, and $M(x)$ is a real analytic function, nonnegative on the support. 
The potential is called {\bf regular} if 
\begin{itemize}
\item [--] the density $\rho$ is strictly positive in the interior of its support.
\item [--] $M$ is takes nonzero values at the endpoints of the support of $\rho$.
\item [--] The inequality in \eqref{ineqs} is strict. 
\end{itemize}
\end{enumerate}
We shall assume regularity and work, without any real loss of generality, near a right endpoint of one of the interval, which will be denoted by $a$. In fact our result is a local result in the neighbourhood of an endpoint such that $M(a_j)>0$ and hence it is a  weaker assumption of regularity that is relevant. 
\bd
The zooming coordinate $\z$ is defined by 
\be
\hbox{(soft edge):}  && \z(z) := \le(\frac 34 n \int_{a}^z M(t) \sqrt{Q(t)} \d t \ri)^{\frac 23}\label{soft} \\
\hbox{(hard edge):} && 
\z(z) := \le(\frac n {4} \int_{0}^z M(t) \sqrt{\frac{Q(t)}t} \d t \ri)^{2} \label{hard}
\ee
and it is a biholomorphic map between a disk centered at  $a$ and a region in the $\z$--plane whose diameter scales as $n^{\frac 23}$ in the soft-edge and $n^2$ in the hard--edge cases, respectively.
\ed
In either case the zooming coordinate has the following expansion
\be
\z(z) =n^\frac 2 3  C (z-a)\le(1 + \mathcal O(z-a)\ri) \qquad \hbox{ or } \qquad
\z(z) =n^2  C z\, \le(1 + \mathcal O(z)\ri),
\ee
where the constant $C$ has the form (respectively)
\be
\label{zoomconst}
C= \le(\frac 3 4 M(a) \sqrt{|Q'(a)|} \ri)^\frac 23 >0\qquad 
\hbox{ or } \qquad
C= \le(\frac {M(0) \sqrt{|Q(0)|}}4\ri)^2>0.
\ee

The scaling limit involves choosing points in the $z$ plane that have a pre-determinate  limit (as $n\to\infty$) in the $\z$ plane;
specifically,  we shall choose points of the form 
\be
&\ds z_j  = a + \frac{\z_j}  {C \,n^{\frac 23 }} \ \ \ &\Rightarrow \z (z) = \z_j + \mathcal O(n^{-2/3})\qquad  \hbox {(soft edge)}\\
&\ds z_j = \frac{\z_j}  {C \,n^{2 }} \ \ \ &\Rightarrow \z (z) = \z_j + \mathcal O(n^{-2}) \qquad \hbox {(hard edge)}.
\ee

The results of \cite{DKMVZ} and \cite{VanlessenUniverBessel} are now summarized in the following two propositions. 
In either case the main object of interest is the matrix kernel 
\be
\mathbb K_n(z_1,z_2)&\&:= \frac{\mu_n(z_1)^{-\frac { \s_3}2 } Y_n^{-1}(z_1) Y_n(z_2)\mu_n(z_1)^{\frac { \s_3}2 }}{z_1-z_2}.
\ee
If one or both the points $z_{1,2}$ belong to $J$ ($\R$ or $\R_+$) then the appropriate indication of the boundary value must be supplied. 
\bp[Soft edge scaling]
\label{softkernel}
Let $z_j = a + \frac {\z_j}{C n^{\frac 2 3}},\ \ j=1,2$. Then 
\be
\lim_{n\to\infty} C n^\frac 2 3 \mathbb K_n(z_1,z_2)=
\frac { \mathcal A^{-1}(\z_1) \mathcal A(\z_2) }  {\z_1-\z_2},
\ee
where $\mathcal A$ is the Airy matrix parametrix as in the Definition \ref{Airyparametrix} below (see \cite{DKMVZ} and references therein), and $C$ is defined in \eqref{zoomconst}.
\ep
\bd 
\label{Airyparametrix}
Let  $\omega:= {\rm e}^{2i\pi/3}$ and $\Ai_j(\z):= \Ai(\omega^j \z)$. The Airy parametrix is defined as
\begin{equation}\label{defA}
   \mathcal A(\zeta)=
   \sqrt{2\pi}e^{-\frac{\pi i}{12}} \times 
   \le\{
   \begin{array}{ll}
        \le[\begin{array}{cc} 
            \Ai(\zeta) & \Ai_2(\zeta) \\
            \Ai'(\zeta) & \Ai_2'(\zeta)
        \end{array}\ri] {{\rm e}^{\frac {-i\pi }{6} \s_3}}, &  \mbox{$\mathrm{for}\; \Im \z>0$,}
\\[15pt]
        \le[\begin{array}{cc} 
            \Ai(\zeta) &-\omega^2 \Ai_1(\zeta) \\
            \Ai'(\zeta) & -\omega^2\Ai_1'(\zeta)
        \end{array}\ri] {{\rm e}^{\frac {-i\pi }{6} \s_3}}, &  \mbox{$\mathrm{for}\; \Im \z < 0$.}
     \end{array}
     \ri.
\end{equation}
\ed
\br\label{Airykernel}
In particular, the $(2,1)$ entry of $\frac { \mathcal A^{-1}(\z_1) \mathcal A(\z_2) }  {\z_1-\z_2}$
\be
\le(\frac { \mathcal A^{-1}(\z_1) \mathcal A(\z_2) }  {\z_1-\z_2}\ri)_{21} &\&= -2i\pi  \ds\frac {
\Ai(\z_1)\Ai'(\z_2)- \Ai'(\z_1) \Ai(\z_2).
}{\z_1-\z_2}
\ee
does not have any jumps on the complex plane and it is the famed Airy kernel (up to the factor $-2i\pi$). This is the limit of the Christoffel--Darboux kernel \eqref{CDkernel}. %
\er
\br
The reader with experience in the nonlinear steepest descent analysis of the Riemann--Hilbert problem \ref{RHPOP} will recognize that the expression for the Airy parametrix above is "simplified" relative to the canon; this is so because we shall not consider the boundary values for the discontinuous entries and hence we can assume that we are (in the relevant cases) outside of the so--called "lenses". The same remark applies to the expression below for the Bessel parametrix.
\er
\bp[Hard edge scaling, see \cite{VanlessenUniverBessel}]
\label{hardkernel}
Let $z_j =  \frac {\z_j}{C n^2},\ \ j=1,2$; then 
\be
\lim_{n\to\infty} C n^2  \mathbb K_n(z_1,z_2)=
\frac { \mathcal B_\nu^{-1}(\z_1) \mathcal B_\nu(\z_2) }  {\z_1-\z_2},
\ee
where $\mathcal B_\nu$ is the Bessel matrix parametrix as in the Definition \ref{Besselparametrix} below, and $C$ is defined in \eqref{zoomconst}.
\ep
\bd
\label{Besselparametrix}
The Bessel parametrix $\mathcal B_\nu(\z)$ is defined as \footnote{In \cite{VanlessenUniverBessel} the parametrix is written for a {\em right} hard edge (and with a slightly different left normalization, which is irrelevant), but here we need a {\em left} hard edge, meaning that the support is to he right of the endpoint. To adapt the formul\ae\ we had to send $\z\to {\rm e}^{i\pi} \z$ and then use some standard identities as in \cite{DMLF} (10.27.6), (10.27.7).}
\be
    \label{BRHPsol}
    \mathcal B_\nu(\zeta)=
     \le[   \begin{array}{cc}
        J_\nu (2\zeta^{\frac{1}{2}}) &
                \frac 1  2 H_\nu^{(1)}  (2\zeta^{\frac{1}{2}}) \\[1ex]
             - 2\pi i  \zeta^{\frac{1}{2}}J_\nu '(2\zeta^{\frac{1}{2}}) &
                -  i\pi \zeta^{\frac{1}{2}}H_\nu^{(1) '}(2\zeta^{\frac{1}{2}})
        \end{array} \ri]
         & 
\ee
where  $J_\nu, H^{(2)}_\nu$ denote the  Bessel and Hankel functions and the square root  in the argument of the functions is taken for $\arg\z\in [0,2\pi)$ (i.e. with the cut along the positive real axis).
\ed
\br
Using the formulas in the electronic library \cite{DMLF}, on can prove that the Bessel functions above satisfy the following boundary conditions on $\R_+$: 
\be\nonumber
&\& H_{\nu}^{(1)}(2\sqrt{\z}_-) 
 \mathop{=}^{\hbox {\tiny  (10.11.5)}}
-{\rm e}^{-i\pi \nu} H_\nu^{(2)}(2\sqrt{\z}_+)
 \mathop{=}^{\hbox {\tiny  (10.4.4)}}
{\rm e}^{-i\pi \nu} (H_\nu^{(1)}(2\sqrt{\z}_+) - 2 J_\nu(2\sqrt{\z}_+))\\ \nonumber
&\& J_\nu(2\sqrt{\z}_-) 
 \mathop{=}^{\hbox {\tiny  (10.11.1)}} {\rm e}^{i\pi \nu}J_\nu( 2\sqrt{\z}_+).
\ee
 Hence we obtain 
 \bea
 \mathcal B_\nu(\z)_+ &=& 
 \mathcal B_\nu(\z)_-\le[
 \begin{array}{cc}
  {\rm e}^{-i\pi \nu} & {\rm e}^{-i\pi\nu}\\
  0 & {\rm e}^{i\pi\nu}
 \end{array}
 \ri],\\
\det \mathcal B_\nu &=& i\pi\sqrt{\z} \le( - J_\nu (2\sqrt \z) H_{\nu}^{(1)'}(2\sqrt{\z}) + J_\nu' (2\sqrt \z) H_{\nu}^{(1)}(2\sqrt{\z}) \ri) 
\mathop{=}^{\hbox{\tiny (10.5.3)}} 1. 
\eea

\er
\br\label{Besselkernel}
In particular, the $(2,1)$ entry of $\frac { \mathcal B_\nu^{-1}(\z_1) \mathcal B_\nu(\z_2) }  {\z_1-\z_2}$ does not have any jumps on the complex plane and it is given by
\be
	\left(\frac { \mathcal B_\nu^{-1}(\z_1) \mathcal B_\nu(\z_2) }  {\z_1-\z_2}\right)_{21} = -2 \pi i \frac{J_\nu(2 \sqrt \z_1)(2\sqrt \z_2)J'_\nu(2\sqrt \z_2) - J_\nu(2 \sqrt \z_2)(2\sqrt \z_1)J'_\nu(2\sqrt \z_1)}{2(\z_1 - \z_2)}.
\ee
which is (up to a rescaling of the variables and the overall constant $-2 \pi i$) the well known Bessel kernel. This is the limit of the Christoffel--Darboux kernel \eqref{CDkernel} near the hard--edge.
\er
Just by juxtaposing  Propositions \ref{finitecorr}, \ref{softkernel} and \ref{hardkernel} we find 
\bc[Soft edge universality]
\label{corsoft}
Let $a$ be a right endpoint of the support of the equilibrium measure, $b_j = a+ \frac{ \beta_j}{Cn^\frac 2 3}$ and $c_j = a+ \frac{ \gamma_j}{Cn^\frac 2 3}$, $j=1\dots S$. Then
\bea
\label{softI}
{\bf [I]} &\& \frac{C^{S^2 } n^{\frac {2S^2}3}}{\ds\prod_{\ell=n}^{n+S-1} h_\ell}  \le\langle  \prod_{j=1}^S  {\rm e}^{-\frac n 2V(b_j) -\frac n 2 V(c_j)} {\det (b_j-M)}{\det(c_j-M)} \ri \rangle = \nonumber \\
&\& = \frac {\le(1 + \mathcal O(n^{-1})\ri)}{ (-2i\pi)^S\Delta(\vec \gamma) \Delta(\vec \beta)} 
\det \le[
\frac {\big(\mathcal A^{-1} (\gamma_\ell ) \mathcal A(\beta_j)\big)_{21}}{\gamma_\ell-\beta_j} 
\ri]_{ \ell, j=1}^S,
\\[15pt]
{\bf [II]}&\&\ \ 
\label{softII}
\le\langle  \prod_{j=1}^S{\rm e}^{-\frac n 2V(b_j)+\frac n2 V(c_j)}  \frac {\det (b_j-M)}{\det(c_j-M)} \ri \rangle = \nonumber\\
&\& =  \frac{\le(1 + \mathcal O(n^{-1})\ri)\prod_{j,\ell} (\gamma_\ell- \beta_j)}{\Delta(\vec \gamma)\Delta(\vec \beta)}
\det \le[
\frac {\big(\mathcal A^{-1}  (\gamma_\ell ) \mathcal A(\beta_j)\big)_{11}}{\gamma_\ell-\beta_j} 
\ri]_{ \ell, j=1}^S,
\\[15pt]
{\bf [III]}&\&\label{softIII}
 \ \ 
\le(\prod_{\ell=n-S}^{n-1} h_\ell\ri)C^{S^2 } n^{\frac {2S^2}3}  \le\langle  \prod_{j=1}^S  \frac 
{{\rm e}^{\frac  n2 V(b_j) + \frac  n 2V(c_j)}} {\det (b_j-M) \det(c_j-M)} \ri \rangle =\nonumber \\
&\&=  
\frac {\le(1 + \mathcal O(n^{-1})\ri)(-2i\pi)^S}{\Delta (\vec \gamma)\Delta (\vec \beta)}
\det \le[
\frac {\big(\mathcal A^{-1}  (\gamma_\ell ) \mathcal A(\beta_j)\big)_{12}}{\gamma_\ell-\beta_j} 
\ri]_{ \ell, j=1}^S.
\eea

\ec

\bc[Hard edge universality]
\label{corhard}
Let $b_j =  \frac{ \beta_j}{Cn^ 2 }$ and $c_j =  \frac{ \gamma_j}{Cn ^2 }$, $j=1\dots S$. Then 
\bea
\label{hardI}
{\bf [I]}&\&\ \frac{C^{S^2 } n^{ {2S^2}}}{\ds\prod_{\ell=n}^{n+S-1} h_\ell}  \le\langle  \prod_{j=1}^S  (b_jc_j)^{\frac \nu 2}  {\rm e}^{-\frac n 2 V(b_j) -\frac n 2 V(c_j)} {\det (b_j-M)}{\det(c_j-M)} \ri \rangle = \nonumber \\
&\& \qquad= \frac {  \le(1 + \mathcal O(n^{-1})\ri)}{ (-2i\pi)^S\Delta(\vec \gamma) \Delta(\vec \beta)} 
\det \le[
\frac {\big(\mathcal B_\nu^{-1} (\gamma_\ell ) \mathcal B_\nu(\beta_j)\big)_{21}}{\gamma_\ell-\beta_j} 
\ri]_{\ell, j=1}^S,
\\
\label{hardII} 
{\bf [II]}&\&\ \ 
\le\langle  \prod_{j=1}^S \le(\frac {b_j}{c_j}\ri)^{\frac \nu 2 }{\rm e}^{-\frac n 2 V(b_j)+\frac n 2V(c_j)}  \frac {\det (b_j-M)}{\det(c_j-M)} \ri \rangle = \nonumber \\
&\& \qquad =\frac{\le(1 + \mathcal O(n^{-1})\ri)\prod_{j,\ell} (\gamma_\ell- \beta_j)}{\Delta(\vec \gamma)\Delta(\vec \beta)}
\det \le[
\frac {\big(\mathcal B_\nu^{-1}  (\gamma_\ell ) \mathcal B_\nu(\beta_j)\big)_{11}}{\gamma_\ell-\beta_j} 
\ri]_{\ell, j=1}^S,
\\[15pt]
{\bf [III]}&\& \ \ 
\label{hardIII}\nonumber 
\le(\prod_{\ell=n-S}^{n-1} h_\ell \ri)C^{S^2 } n^{{2S^2}}  \le\langle  \prod_{j=1}^S  \frac {(c_j b_j)^{-\frac \nu 2}{\rm e}^{\frac n 2 V(b_j) + \frac n 2 V(c_j)}} {\det (b_j-M) \det(c_j-M)} \ri \rangle =\\
&\&\qquad=  
\frac {\le(1 + \mathcal O(n^{-1})\ri)(-2i\pi)^S}{\Delta (\vec \gamma)\Delta (\vec \beta)}
\det \le[
\frac {\big(\mathcal B_\nu^{-1}  (\gamma_\ell ) \mathcal B_\nu (\beta_j)\big)_{12}}{\gamma_\ell-\beta_j} 
\ri]_{\ell, j=1}^S.
\eea
\ec
\br
For the one-cut case  $[a,b]$ (soft-edge) or $(0,b]$ (hard-edge) the constants $\le(\prod_{\ell=n-S}^{n-1} h_\ell \ri)^{\mp 1}$ appearing in the left hand side of \eqref{softI},\eqref{hardI} and \eqref{softIII},\eqref{hardIII} respectively have a simple expression because 
\be
h_{n+r} = \frac \pi 2 {\rm e}^{n\ell}(b-a) (1 + \mathcal O(n^{-1})), \ \ \ r\in \Z \mbox{ bounded}.
\ee
and hence
\be
\le(\prod_{\ell=n-S}^{n-1} h_\ell \ri)^{\mp 1}  =\frac{\pi^{\mp S} {\rm e}^{\mp nS\ell}(b-a)^{\mp S}}{2^{\mp S}} (1+ \mathcal O(n^{-1})).
\ee

If the support of the equilibrium measure consists of two or more intervals, the asymptotic behaviour is of the form  $h_{n+r} = {\rm e}^{n \ell} F_{n,r}$, where $F_{n,r}$ is a rather complicated expression (but uniformly bounded in $n$)  in terms of Riemann theta functions.
\er

\section{The matrix Airy and Bessel functions}
The proof of the two Theorem \ref{thmsoft}, \ref{thmhard} now requires only to identify the right sides of Corollaries \ref{corsoft} and \ref{corhard} with the appropriate expressions in terms of matrix integrals. 
We need the following determinantal identity.
\bl
\label{LemmaDet}
Let $H_{j,k} = \le[\begin{array}{cc}
a_j & b_j\\
c_k & d_k
\end{array}\ri]$ be matrices in terms of the indicated entries, $j,k=1\dots, S$,   and $x_1\dots x_S, y_1,\dots y_S$ arbitrary (complex) numbers.
Then 
\be
\label{detid}
\det\le[\frac {\det H_{j,k}}{x_j-y_k} \ri]_{1\leq j,k\leq S}
=\frac{ (-1)^{S(S-1)/2}}{\prod_{j,k}(x_j-y_k)}
\det \le[
\begin{array}{c|c}
\le[x_j^{\ell-1} a_j \ri]_{j,\ell \leq S}
 & 
\le[x_j^{\ell-1} b_j
\ri]_{j,\ell\leq S}
\\[18pt]
\hline\\
\le[ y_j^{\ell-1} c_j\ri]_{j,\ell \leq S} & 
\le[ y_j^{\ell-1} d_j \ri]_{j,\ell\leq S}
\end{array}
\ri]
\ee
(in the blocks above, $j$ is the row-index and $\ell$ is the column).
\el
{\bf Proof.}
This identity is deduced easily as a special case of  Thm. 1.1(a) in \cite{IshikawaOkada}.
With reference to loc. cit, we use $p=q=0, n=S\in \N$ and then their identity reads (the index $j$ is the row-index, $\ell$ the column index)
\be
\det \le[
\frac {\det \le[
\begin{array}{cc}
1 & q_j\\
1 & p_k
\end{array}
\ri]}{x_j-y_k}
\ri]_{j,k=1}^S
= \frac {(-1)^{S(S-1)/2}}{\prod_{j,k}(x_j-y_k)} \det 
\le[
\begin{array}{c|c}
\le[x_j^{\ell-1}  \ri]_{j,\ell \leq S}
 & 
\le[x_j^{\ell-1} q_j
\ri]_{j,\ell\leq S}
\\[18pt]
\hline\\
\le[ y_j^{\ell-1}\ri]_{j,\ell \leq S} & 
\le[ y_j^{\ell-1}p_j \ri]_{j,\ell\leq S}
\end{array}
\ri]
\label{Ishikawa}
\ee
The identities that we need have a similar form; on the right side of \eqref{detid}
we factor out from the first $n$ rows the determinant of ${\rm diag}(a_1,\dots, a_S)$ and from the second block the determinant of ${\rm diag}(c_1,\dots, c_S)$ reduces the determinant to the same as in \eqref{Ishikawa} with $q_j = \frac {b_j}{a_j}$ and $p_j= \frac {d_j}{c_j}$. Of course we assume all $a_j, c_j$ to be nonzero. 
Then the identity is easily established; the case where some $a_j$ or $c_j$ are zero  holds by analytic continuation and the obvious  homogeneity.
\QED 
Now, applying the Lemma above to the expressions appearing in Corollaries \ref{corsoft} and \ref{corhard}, we obtain the following alternative expressions. Below $\mathcal P = \mathcal A$ or $\mathcal B_\nu$ depending on the case (hard/soft edge).
\bea
{\bf [I]}\qquad  \frac {1}{ \Delta(\vec \gamma) \Delta(\vec \beta)} 
\det \le[
\frac {\big(\mathcal P^{-1} (\gamma_\ell ) \mathcal P(\beta_j)\big)_{21}}{(-2i\pi)(\gamma_\ell-\beta_j)} 
\ri]_{ \ell, j=1}^S \hspace{-20pt}
&\&=(-1)^{\frac S2(S-1)} 
\frac {\det \le[
\begin{array}{c|c}
\le[\gamma_j^{\ell-1} \mathcal P_{11}(\gamma_j) \ri]_{j,\ell \leq S}
 & 
\le[\gamma_j^{\ell-1} \mathcal P_{21}(\gamma_j) 
\ri]_{j,\ell\leq S}
\\[18pt]
\hline\\
\le[\beta_j^{\ell-1} \mathcal P_{11}(\beta_j) \ri]_{j,\ell \leq S}
 & 
\le[\beta_j^{\ell-1} \mathcal P_{21}(\beta_j) 
\ri]_{j,\ell\leq S}\end{array}
\ri]}{(-2i\pi)^S\Delta (\vec \gamma) \Delta (\vec \beta) \prod_{j,k}(\gamma_j-\beta_k)}\nonumber\\ \label{altsoftI}
\eea
\bea
{\bf [II]}\qquad  \frac {1}{ \Delta(\vec \gamma) \Delta(\vec \beta)} 
\det \le[
\frac {\big(\mathcal P^{-1} (\gamma_\ell ) \mathcal P(\beta_j)\big)_{11}}{(-2i\pi)(\gamma_\ell-\beta_j)} 
\ri]_{ \ell, j=1}^S \hspace{-20pt}
&\&=(-1)^{\frac S2(S-1)} 
\frac {\det \le[
\begin{array}{c|c}
\le[\gamma_j^{\ell-1} \mathcal P_{22}(\gamma_j) \ri]_{j,\ell \leq S}
 & 
\le[\gamma_j^{\ell-1} \mathcal P_{21}(\gamma_j) 
\ri]_{j,\ell\leq S}
\\[18pt]
\hline\\
\le[\beta_j^{\ell-1} \mathcal P_{12}(\beta_j) \ri]_{j,\ell \leq S}
 & 
\le[\beta_j^{\ell-1} \mathcal P_{11}(\beta_j) 
\ri]_{j,\ell\leq S}\end{array}
\ri]}{(-2i\pi)^S\Delta (\vec \gamma) \Delta (\vec \beta) \prod_{j,k}(\gamma_j-\beta_k)}\nonumber\\ \label{altsoftII}
\eea
\bea
{\bf [III]}\qquad  \frac {1}{ \Delta(\vec \gamma) \Delta(\vec \beta)} 
\det \le[
\frac {\big(\mathcal P^{-1} (\gamma_\ell ) \mathcal P(\beta_j)\big)_{12}}{(-2i\pi)(\gamma_\ell-\beta_j)} 
\ri]_{ \ell, j=1}^S \hspace{-20pt}
&\&
=
(-1)^{\frac S2(S-1)} 
\frac {\det \le[
\begin{array}{c|c}
\le[\gamma_j^{\ell-1} \mathcal P_{22}(\gamma_j) \ri]_{j,\ell \leq S}
 & 
\le[\gamma_j^{\ell-1} \mathcal P_{12}(\gamma_j) 
\ri]_{j,\ell\leq S}
\\[18pt]
\hline\\
\le[\beta_j^{\ell-1} \mathcal P_{22}(\beta_j) \ri]_{j,\ell \leq S}
 & 
\le[\beta_j^{\ell-1} \mathcal P_{12}(\beta_j) 
\ri]_{j,\ell\leq S}\end{array}
\ri]}{(-2i\pi)^S\Delta (\vec \gamma) \Delta (\vec \beta) \prod_{j,k}(\gamma_j-\beta_k)}\nonumber\\ \label{altsoftIII}
\eea

\subsection{Proof of Theorem \ref{thmsoft}}

To prove the Theorem \ref{thmsoft} we finally combine the Corollary \ref{corsoft}, equation \eqref{softI}, with the equation \eqref{altsoftI} and we perform some elementary algebraic manipulation on the right hand side of  the latter. Let us denote with $\vec x = (\vec \gamma,\vec \beta)$ the collection of all the $2S$ points in \eqref{altsoftI}.
Recalling that, by definition, $\mathcal A_{11}(\z) = \sqrt{2\pi} {\rm e}^{-(\pi i)/4}\Ai(\z)$ and that $\mathcal A_{21}$ is its derivative, we obtain 
\bea
&\& (-1)^{\frac S2(S-1)} 
\frac {\det \le[
\begin{array}{c|c}
\le[\gamma_j^{\ell-1} \mathcal A_{11}(\gamma_j) \ri]_{j,\ell \leq S}
 & 
\le[\gamma_j^{\ell-1} \mathcal A_{21}(\gamma_j) 
\ri]_{j,\ell\leq S} \nonumber
\\[18pt]
\hline\\
\le[\beta_j^{\ell-1} \mathcal A_{11}(\beta_j) \ri]_{j,\ell \leq S}
 & 
\le[\beta_j^{\ell-1} \mathcal A_{21}(\beta_j) 
\ri]_{j,\ell\leq S}\end{array}
\ri]}{(-2i\pi)^S\Delta (\vec \gamma) \Delta (\vec \beta) \prod_{j,k}(\gamma_j-\beta_k)} = \\ \nonumber
&\& \nonumber \\ \nonumber
&\&  = \frac{(-1)^{\frac S2(S-1)} (\sqrt{2 \pi} {\rm e}^{-\frac {\pi i}4})^{2S}}{(-2\pi i)^{S} \Delta(\vec x)} \det 
\le[\begin{array}{c c c | c c c} 
		\Ai(x_1) & \ldots & x_1^{S-1}\Ai(x_1) & \Ai'(x_1) & \ldots & x_1^{S-1}\Ai'(x_1)\\
		\Ai(x_2) & \ldots & x_2^{S-1}\Ai(x_2) & \Ai'(x_2) & \ldots & x_2^{S-1}\Ai'(x_2)\\
		\vdots & \vdots & \vdots & \vdots & \vdots & \vdots \\
		\Ai(x_{2S}) & \ldots & x_{2S}^{S-1}\Ai(x_{2S}) & \Ai'(x_{2S}) & \ldots & x_{2S}^{S-1}\Ai'(x_{2S})
\end{array}\ri] =\\
&\&\nonumber \\ \nonumber
&\&  = \frac{ 1}{\Delta(\vec x)} 
\le[\begin{array}{cccccc}
		\Ai(x_1) & \ldots & x_1^{S-1}\Ai(x_1) & \Ai'(x_1) & \ldots & x_1^{S-1}\Ai'(x_1) \\
		\Ai(x_2) & \ldots & x_2^{S-1}\Ai(x_2) & \Ai'(x_2) & \ldots & x_2^{S-1}\Ai'(x_2) \\
		\vdots & \vdots & \vdots & \vdots & \vdots & \vdots \\
		\Ai(x_{2S}) & \ldots & x_{2S}^{S-1}\Ai(x_{2S}) & \Ai'(x_{2S}) & \ldots & x_{2S}^{S-1}\Ai'(x_{2S}) 
\end{array}\ri] = \\
&\&  = \frac{\det \le[\Ai^{(\ell-1)}(x_j)\ri]_{\ell,j=1}^{2S}}{\Delta(\vec x)}. \nonumber
\eea
where, in the last passage, we used the differential equation $\Ai''(\z) = \z\Ai(\z)$. On the other hand, it is a well known result that the (even dimensional) matrix Airy function \cite{Kontsevich} 
\bea
\label{ZnK}
Z_{2S}({Y}):=\frac{ \ds \int_{\mathcal H_{2S}} \d H {\rm e}^{ \Tr \le(i\frac {H^3}3 - {Y} H^2 \ri)}}{ \int_{H_{2S}} \d  H {\rm e}^{- \Tr \le( {Y} H^2\ri)}}.
\eea
satisfies the following equation (see loc. cit. or Appendix B in \cite{BertolaCafassoKonts}):
 \be
 \label{ZnKY}
 Z_{2S}({Y}) = 2^{2S} \pi^S {\rm e}^{\frac 2 3 \Tr  {Y}^3}\frac{\det\le[\Ai^{(j-1)} (y_k^2 ) \ri]_{k,j=1}^{2S} \prod_{j=1}^{2S}(y_j)^{\frac 1 2} }{\prod_{j<k} (y_j-y_k) }, \ \ \ \Re y_j>0.
 \ee
 Then the proof of the Theorem \ref{thmsoft} proceeds by simple algebra identifying the squares of the variables $y_j$'s with the $x_j$'s.
\br\label{RemarkAiry}
En passant, our computations (see Remark \ref{Airykernel}) show the remarkable identity
\be
	\frac{\det\Big(K_\Ai (\gamma_\ell,\beta_j)\Big)_{j,l = 1}^S}{\Delta(\vec \gamma)\Delta(\vec \beta)} = \frac{{\rm e}^{- \frac{2}3 \Tr(Y^3)}\prod_{j < k}(y_j + y_k)}{2^{2S}\pi^S\prod \sqrt y_j}Z_{2S}(Y)
\ee
 under the identification $(\gamma_1,\ldots\gamma_S,\beta_1,\ldots,\beta_S) = (y_1^2,\ldots,y_{2S}^2)$, see also \cite{OkounkovKdV, ASvM}.
\er
\br \label{generalcase}
In the case $III$ for the soft edge, the expression \eqref{altsoftIII} in general is not reducible directly to a matrix integral, unless all points are in the upper or lower half-plane, in which case the expression is the same (up to un-interesting overall constants) with the replacement $\Ai\mapsto \Ai_{1,2}$, (where $\Ai_{j}(\z):= \Ai(\omega^j \z),\ \omega:= {\rm e}^{\frac {2i\pi}3}$) depending on the case according to \eqref{Airyparametrix}. On the other hand, it is known that the Kontsevich matrix model \eqref{ZnKY} admits a regular asymptotic expansion at infinity just when all the points $\{y_j, j =1,\ldots,2S\}$ have positive real part. If one wants to extend \eqref{ZnKY} to a function admitting regular expansion at infinity for arbitrary $y_j$'s, it is necessary to introduce a \emph{generalized Kontsevich matrix model} 
of the form
\be
\label{ZnKcont}
Z_n(\mathcal Y^{(0)}, \mathcal Y^{(1)}, \mathcal Y^{(2)} ) =
(-\omega)^{n_1 -n_2}
( 2\sqrt \pi)^{n}\frac{ {\rm e}^{\frac 23  \Tr Y^3  + x \Tr Y} \prod_{j=1}^{n} (y_j)^\frac 1 2 }{\prod_{j<k}(y_j-y_k)}\det \le[
\begin{array}{cc}
\ds \le[ \Ai_0^{(k-1)} ( y_j^2 + x)\ri]_{y_j\in \mathcal Y^{(0)}\atop 1\leq k \leq n} \\
\hline 
\ds\le[ \Ai_1^{(k-1)}( y_j^2 + x)\ri]_{y_{j}\in \mathcal Y^{(1)}\atop 1 \leq k \leq n}
\\
\hline 
\ds\le[ \Ai_2^{(k-1)}( y_j^2 + x)\ri]_{y_{j}\in \mathcal Y^{(2)}\atop 1 \leq k \leq n}
\end{array}
\ri].
\ee
Here $\omega = {\rm e}^{\frac{2 \pi i }3}$ and the set $\{y_j , j =1, \ldots 2S\}$ is divided into three subsets $\mathcal Y^{(0)}, \mathcal Y^{(1)}$ and  $\mathcal Y^{(2)}$ according to the positions of the $y_j$'s on the complex plane, in such a way that all the entries in the matrix (multiplied by the pre factor ${\rm e}^{\frac 23  y_j^3  + x y_j} (y_j)^\frac 1 2$) admit regular expansion at infinity (for more details, see \cite{BertolaCafassoKonts}, Definition 1.2).\\
Now, going back to \eqref{altsoftIII}, soft edge case, suppose that $(\vec x) = (y_1^2,\ldots,y_{2S}^2)$ is such that the first $S_2$ components lie in the upper half plane, and the rest of them in the lower half plane\footnote{The assumption is made just for simplicity to avoid one more shuffling of lines in the matrices written below.}. Then the right hand side of \eqref{altsoftIII} will give

\bea
&\& \nonumber (-1)^{\frac S2(S-1)} 
\frac {\det \le[
\begin{array}{c|c}
\le[\gamma_j^{\ell-1} \mathcal A_{22}(\gamma_j) \ri]_{j,\ell \leq S}
 & 
\le[\gamma_j^{\ell-1} \mathcal A_{12}(\gamma_j) 
\ri]_{j,\ell\leq S}
\\[18pt]
\hline\\
\le[\beta_j^{\ell-1} \mathcal A_{22}(\beta_j) \ri]_{j,\ell \leq S}
 & 
\le[\beta_j^{\ell-1} \mathcal A_{12}(\beta_j) 
\ri]_{j,\ell\leq S}\end{array}
\ri]}{(-2i\pi)^S\Delta (\vec \gamma) \Delta (\vec \beta) \prod_{j,k}(\gamma_j-\beta_k)} = \\
&\& \nonumber \\
&\&= \frac{(-1)^{\frac S2(S-1)} (\sqrt{2 \pi} {\rm e}^{\frac {\pi i}{12}})^{2S}(-\omega^2)^{S_2}}{(-2\pi i)^{S} \Delta(\vec x)} \det 
\le[\begin{array}{c c c | c c c} 
		\Ai_2'(x_1) & \ldots & x_1^{S-1}\Ai_2'(x_1) & \Ai_2(x_1) & \ldots & x_1^{S-1}\Ai_2(x_1)\\
		\vdots & \vdots & \vdots & \vdots & \vdots & \vdots \\
		\Ai_2'(x_{S_2}) & \ldots & x_{S_2}^{S-1}\Ai'_2(x_{2S}) & \Ai_2(x_{S_2}) & \ldots & x_{S_2}^{S-1}\Ai_2(x_{S_2}) \\
		&&&&&\\
		\hline &&&&&\\
		\Ai_1'(x_{S_2 + 1}) & \ldots & x_{S_2 + 1}^{S-1}\Ai_1'(x_{S_2 + 1}) & \Ai_1(x_{S_2 + 1}) & \ldots & x_{S_2 + 1}^{S-1}\Ai_1(x_{S_2 + 1})\\
		\vdots & \vdots & \vdots & \vdots & \vdots & \vdots \\
		\Ai_1'(x_{2S}) & \ldots & x_{2S}^{S-1}\Ai'_1(x_{2S}) & \Ai_1(x_{2S}) & \ldots & x_{2S}^{S-1}\Ai_1(x_{2S})
\end{array}\ri] =
\nonumber\\
&\&\nonumber \\
&\& \nonumber  = \frac{ {\rm e}^{\frac{\pi i}6}(-\omega^2)^{2S - S_2}}{\Delta(\vec x)} 
\le[\begin{array}{cccccc}
		\Ai_2(x_1) & x_1\Ai_2(x_1) & \Ai'_2(x_1) & x_1\Ai'_2(x_1) & \ldots & x_1^{S-1}\Ai'_2(x_1) \\
		\vdots & \vdots & \vdots & \vdots & \vdots & \vdots \\
		\Ai_2(x_{S_2}) & x_{S_2}\Ai_2(x_{S_2}) & \Ai_2'(x_{S_2}) & x_{S_2}\Ai'_1(x_{S_2}) & \ldots & x_{S_2}^{S-1}\Ai_2'(x_{S_2}) \\
		&&&&&\\
		\hline\\
		\Ai_1(x_{S_2+1}) & x_{S_2+1}\Ai_1(x_{S_2+1}) & \Ai_1'(x_{S_2+1}) & x_{S_2+1}\Ai'_1(x_{S_2+1}) & \ldots & x_{S_2+1}^{S-1}\Ai_1'(x_{S_2+1}) \\
		\vdots & \vdots & \vdots & \vdots & \vdots & \vdots \\
		\Ai_1(x_{2S}) & x_{2S}\Ai_1(x_{2S}) & \Ai_1'(x_{2S}) & x_{2S}\Ai_1'(x_{2S}) & \ldots & x_{2S}^{S-1}\Ai'_1(x_{2S}) 
\end{array}\ri] = \\
&\&  \nonumber = \frac{ {\rm e}^{\frac{\pi i}6}(-\omega^2)^{2S - S_2}}{\Delta(\vec x)}  
	\det\left[ \begin{array}{c}
			\left[ \Ai_2^{k-1}(x_j)\right]_{1 \leq j \leq S_2 \atop 1 \leq k \leq S}\\
			\hline
			\left[ \Ai_1^{k-1}(x_j)\right]_{S_2+1 \leq j \leq 2S \atop 1 \leq k \leq S}
		\end{array} \right].
\eea
The determinant in the last line is of the type \eqref{ZnKcont}, and hence we obtain an analogue of the Theorem \ref{thmsoft} stating that
\be
\lim_{n\to\infty} \le(\prod_{\ell=n-S}^{n-1} h_\ell\ri)C^{S^2 } n^{\frac {2S^2}3}  \le\langle  \prod_{j=1}^{2S}  \frac 
{{\rm e}^{\frac  n2 V(x)}} {\det (x_j-M)} \ri \rangle_{\mathcal H_n}\!\!\!
= \kappa \frac{\ds\prod_{j<k} (y_j+ y_k)}{ \ds\prod_{j = 1}^{2S} \sqrt{y_j}} {\rm e}^{-\frac 23 \Tr Y^3} Z_{2S}^{Kont}({\mathcal Y}^{(1)},{\mathcal Y}^{(2)}),
\label{softdualitygeneralised}
\ee
where $\kappa$ is an unimportant constant, $x_j = a + C^{-1} n^{-\frac 2 3}  y_j^2$ and $\mathcal Y^{(2)} := (y_1,\ldots,y_{S_2})$, $\mathcal Y^{(1)} := (y_{S_2 + 1},\ldots,y_{2S})$.\\
As the reader can imagine, analogous computations show that also \eqref{altsoftII} can be rewritten in terms of the generalized Kontsevich matrix model. 
\er

\subsection{Proof of Thm. \ref{thmhard}}
For the hard-edge case we can write matrix-integral expression in both case $I,III$; the computation is parallel and simplified by noticing first that the second row of $\mathcal B_\nu$ \eqref{BRHPsol} is $-2i\pi \z\frac {\d}{\d \z}$ of the first.
The first row is given in terms of the following integral representations;

\be
\le[\mathcal P_{11}(\z), \mathcal P_{12}(\z)\ri]=\sqrt{2\pi} 
\Bigg[ {\z^{\frac \nu 2 }} \underbrace{\int_{\wh \gamma} {\rm e}^{-\z s + \frac 1  {s}} s^\nu \frac {\d s}{2i\pi s}}_{:=f_\nu(\z)} \ ,\ 
{\z^{\frac \nu 2 }} \underbrace{\int_{0_-}^{\frac {|\z|}{\z}\infty} 
\hspace{-10pt}
{\rm e}^{-\z s +\frac 1  s } s^\nu\frac {\d s}{2i\pi s}}_{:=g_\nu(\z)}\Bigg]
\ee
The power $s^\nu$ in the integrands of $f_\nu, g_\nu$ is defined with a cut along the $s\in \R_-$. The powers of $\z$ are with $\arg\z\in [0,2\pi)$. 

In what follows, we denote $a_\nu := \sqrt{2\pi} \z^\frac \nu 2$.  Using integration by parts and trivial algebra, we find
\be
\label{secondrow}
\mathcal P_{21}(\z) ={2i\pi } a_\nu\le( f_{\nu - 1} -\frac \nu 2  f_\nu\ri),
\qquad
\mathcal P_{22}(\z) ={2i\pi } a_\nu\le( g_{\nu - 1} -\frac \nu 2  g_\nu\ri)
\ee
and both $f_\nu, g_\nu$ satisfy the same recurrence relation
\be
\z f_\nu(\z)  = (\nu-1) f_{\nu-1}(\z)- f_{\nu-2}(\z), \label{recx}\\
\z g_\nu(\z) = (\nu-1) g_{\nu-1}(\z)- g_{\nu-2}(\z). \label{recx2}
\ee

 Now let us fix some notations to simplify the sequel of the proof: first of all, similarly as before, we denote with $\vec y = (y_1,\dots, y_{2n}) := (\vec \gamma,\vec \beta) $ the set of all points $\{\gamma_j, \beta_k\}$. Then we set  $F(\vec y) := (F(y_1),\dots, F(y_{2n}))$. Also we adopt temporarily the ``vectorization'' convention: namely, products of type $\vec a \vec y$ are Hadamard products, i.e. entry-wise.\\
 
With these conventions, the numerator of the right side of \eqref{altsoftI} reads 
\be
&\& (-1)^{\frac S 2(S-1)}
\det\le[
\begin{array}{c}
{\mathcal P}_{11} (\vec y)\\
\vec y\, {\mathcal P}_{11} (\vec y)\\
\vdots
\\
\vec y^{S-1}{\mathcal P}_{11} (\vec y)\\
{\mathcal P}_{21} (\vec y)\\
\vec y\, {\mathcal P}_{21} (\vec y)\\
\vdots\\
\vec y^{S-1}{\mathcal P}_{21} (\vec y)
\end{array}
\ri]=
\det\le[
\begin{array}{c}
{\mathcal P}_{11} (\vec y)\\
{\mathcal P}_{21} (\vec y)\\
\vec y\,{\mathcal P}_{11} (\vec y)\\
\vec y\,{\mathcal P}_{21} (\vec y)\\
\vdots
\\ \nonumber
\vec y^{S-1}{\mathcal P}_{11} (\vec y)\\
\vec y^{S-1}{\mathcal P}_{21} (\vec y)
\end{array}
\ri]
\mathop{=}^{\eqref{secondrow}}\det \le[
\begin{array}{c}
\vec a_\nu f_\nu(\vec y)\\
2i\pi  {\vec a_\nu} \le({f}_{\nu-1} - \frac \nu 2  f_\nu\ri)(\vec y) \\[6pt]
\vec a_\nu \vec y f_\nu(\vec y)\\
 2i\pi {\vec a_\nu}\vec y \le({f}_{\nu-1} - \frac \nu 2 f_\nu\ri)(\vec y) \\[6pt]
\vdots
\\
\vec a_\nu \vec y^S f_\nu(\vec y)\\
2i\pi  {\vec a_\nu} \vec y ^S \le(  f_{\nu-1} - \frac \nu 2  f_\nu\ri)(\vec y) \\
\end{array}
\ri]
=\nonumber \\ \nonumber
&\&\mathop{=}(2i\pi)^{S }\prod_{j=1}^{2S} a_\nu(y_j)
\det \le[
\begin{array}{c}
 f_\nu(\vec y)\\
 \le(f_{\nu-1} - \frac \nu  2 f_\nu\ri)(\vec y) \\
 \vec y f_\nu(\vec y)\\
 \vec y\le(f_{\nu-1} - \frac \nu 2 f_\nu\ri)(\vec y) \\
\vdots
\\
\vec y^S f_\nu(\vec y)\\
 \vec y^S\le(f_{\nu-1} - \frac \nu 2 f_\nu\ri)(\vec y) \\
\end{array}
\ri]
\mathop{=}^{\eqref{recx}}
(-2i\pi)^S\prod_{j=1}^{2S} a_\nu(y_j)
\det \le[
\begin{array}{c}
 f_\nu(\vec y)\\
f_{\nu-1}(\vec y) \\
f_{\nu-2}(\vec y)\\
f_{\nu-3}(\vec y)  \\
 f_{\nu-4}(\vec y)\\
 f_{\nu-5}(\vec y)  \\
\vdots
\\
 f_{\nu-2S+2}(\vec y)\\
f_{\nu-2S+1}(\vec y)  \\
\end{array}
\ri]
=\\
&\&=
(-2i\pi)^S\prod_{j=1}^{2S} a_\nu(y_j)
\det \bigg[
f_{\nu-\ell+1}(y_j)
\bigg]_{1\leq \ell, j\leq 2S}.\label{3.16}
\eea
For the sequel we recall that $Y = \mathrm{diag}(y_1,\ldots,y_{2K})$ With a slight abuse of notation, we will also denote $\Delta(Y)$ the Vandermonde determinant of the same variables. The second chain of identities starts applying the standard Andreief and Harish--Chandra (HC) identities (this latter supplemented with some computations in the Appendix \ref{HCconst}):
\bea
&\& 
\eqref{3.16}
\mathop{=}^{\hbox{\tiny Andreief}}
(2\pi)^S \det ( Y)^{\frac \nu 2} \frac {(-2i\pi)^{S}}{ (2S)!}
\int_{\wh \gamma^{2S}}  \Delta (\vec t) \det\bigg[{\rm e}^{-y_j t_k}\bigg]_{j,k\leq 2n} \prod_{j}{\rm e}^{\frac 1{t_j}}t_j^{\nu-2S+1}\frac{\d t_j}{2i\pi t_j} =\\
&\& \mathop{=}^{\hbox{\tiny HC + App. \ref{HCconst} }}
\frac {(2\pi)^{2S} (-i)^S \det ({Y})^\frac \nu  2\Delta (Y)}{ \pi^{S(2S-1)} }
\int_{\wh \gamma^{2S}} \Delta (\vec t)^2  \int_{U(2S)}\hspace{-10pt} {\rm e}^{-\Tr(T U Y U^\dagger)}  \d U \, \prod_{j}{\rm e}^{\frac 1{t_j}}t_j^{\nu-2S+1}\frac{\d t_j}{2i\pi t_j}
\mathop{=}\label{417} \\
&\& =\label{418}
\frac {(-i)^{-S} \det ( {Y})^\frac \nu  2\Delta (Y)}{  \pi^{S(2S-1)}}
\int_{\wh \gamma^{2n}} \Delta (\vec t)\Delta \le(\frac 1{\vec t}\ri) \prod_{j=1}^{2S} t_j^{2S-1} \int_{U(2S)}\hspace{-12pt} {\rm e}^{-\Tr(T U Y U^\dagger)}  \prod_{j}{\rm e}^{\frac 1{t_j}}t_j^{\nu-2S+1}\frac{\d t_j}{t_j}
\mathop{=} \\
&\& =
\frac { (-i)^{-S} \det ( {Y})^\frac \nu  2\Delta (Y)}{ \pi^{S(2S-1)}}
\int_{CUE_{2S,\wh \gamma}} (\det M)^{\nu-1}
\exp \le[\Tr\le(-Y M + M^{-1}\ri)\ri] \d M.
\ee
In the step from \eqref{417} to \eqref{418} we have factored $\Delta(\vec t) = \Delta \le(\frac 1{\vec t}\ri) \prod_{j=1}^{2S} t_j^{2S-1}$ because, in the case of the CUE proper, the Jacobian of the change of variables from angular to eigenvalues is $\Delta(\vec t) \ov {\Delta(\vec t)}$ and $\ov t_j =\frac 1 {t_j}$. 
Indeed, 
if $\nu$ is an integer, the contour of integration $\wh \gamma$ can be chosen as the unit circle and we get a bona fide CUE.

\br\label{RemarkBessel}
 Analogously as in the case of the soft edge, taking into account the Remark \ref{Besselkernel} we have obtained the interesting identity 
\be
\frac{\det\le[K_{B}(\beta_j,\gamma_k)\ri]}{\Delta (\vec \beta)\Delta (\vec \gamma)}  = 
\frac{\det(Y)^{\frac{\nu}2}(2 \pi)^S}{\pi^{S(2S - 1)}}\int_{CUE_{2S,\wh \gamma}} (\det M)^{\nu-1}
\exp \le[\Tr\le(-Y M + M^{-1}\ri)\ri] \frac{\d M}{(2i\pi)^{2S}}
\label{detBessel}
\ee
where $K_B$ is the Bessel kernel. 
\er

For the case $III$ the computation is essentially  identical, only replacing $f_\nu \mapsto g_\nu$: since the support of the measure is on $\R_+$, it is more meaningful to take the points away from $\R_+$; 
 specifically we take $-\vec y = (\vec \beta,\vec \gamma)$, with $ \Re \beta_j<0$, $\Re \gamma_i < 0 $ so that $\Re y_\ell>0$. With this proviso we need the evaluation of the functions $\mathcal P_{12}(-\xi)$, $\Re \xi>0$ which can be 
written  
\be
\mathcal P_{12}({\rm e}^{i\pi}\xi) = {\rm e}^{ \frac {i \pi} 2 \nu}\xi^\frac \nu 2 g_\nu({\rm e}^{i\pi} \xi)\ ;\qquad 
g_\nu({\rm e}^{i\pi} \xi) ={\rm e}^{-i\pi \nu}  \int_{\R_+}  {\rm e}^{-\xi s - s} s^\nu \frac {\d s}{2i\pi s}.
\ee

and,  in view of the slightly different normalization in \eqref{hardIII} we obtain
\bea
{\bf [III]}&\& \ \ 
\nonumber 
\lim_{n\to\infty}\le(\prod_{\ell=n-S}^{n-1} h_\ell \ri)C^{S^2 } n^{{2S^2}}  \le\langle  \prod_{j=1}^S  \frac {(c_j b_j)^{-\frac \nu 2}{\rm e}^{\frac n 2 V(b_j) + \frac n 2 V(c_j)}} {\det (b_j-M) \det(c_j-M)} \ri \rangle_{\mathcal H_n^+} =\\ \nonumber
&\&\qquad=  
\frac {(-2i\pi)^S}{\Delta (\vec \gamma)\Delta (\vec \beta)}
\det \le[
\frac {\big({\mathcal P}^{-1}  (\gamma_\ell ) {\mathcal P} (\beta_j)\big)_{12}}{\gamma_\ell-\beta_j} 
\ri]_{\ell, j=1}^S\hspace{-20pt}
=\\
&\&=\frac { 2^S  {\rm e}^{ -i\pi S \nu }\det ( {Y})^\frac \nu  2}{  \pi^{2S(S-1)}}
\int_{\mathcal H^+_{2S}} \det ( M)^{\nu-2S}
\exp \le[\Tr\le(-Y M - M^{-1}\ri)\ri] {\d M}
\label{detBesselK}
\eea
The different power of the determinant in the integrand of \eqref{detBesselK} is due to the different meaning of the symbol $\d M$: in this case it means the Lebesgue measure and, $\d M \propto \d U \prod  \d t_j \Delta(\vec t)^2$ while in the $CUE_{\wh \gamma}$  case $\d M \propto \d U \prod  \d t_j \Delta(\vec t)\Delta (\frac 1 {\vec t})$.

\paragraph{Acknowledgements.} 
The research of M.B  was supported in part by the Natural Sciences and Engineering Research Council of Canada grant
RGPIN/261229--2011 and by the FQRNT grant "Matrices Al\'eatoires, Processus Stochastiques et Syst\`emes Int\'egrables" (2013--PR--166790).
The research of M.C. was partially supported by a project ``Nouvelle \'equipe'' funded by the region Pays de la Loire. M.C. thanks the International School of Advanced Studies (SISSA) in Trieste for the hospitality during the preparation of this work.
 
 \appendix

\section{Normalization constants for the Harish-Chandra formula}
\label{HCconst}
In the proof of the Theorem \ref{thmhard}, we used the Harish-Chandra formula
{stating that, given a unitary ensemble with potential $V$ and external potential $\Lambda$,} 
\be\label{HCAppendix}
\int_{\mathcal H_n} {\rm e}^{-\Tr (V(M) - M \Lambda)} \d M = \frac {K_n}{\Delta (\Lambda)} \int_{\R_n} \Delta(X) \det \le[{\rm e}^{x_j \l_k}\ri]_{j,k=1}^{n} \prod_{\ell=1}^n {\rm e}^{-V(x_j)} \d x_j,
\ee
where the constant $K_n$ depends on $n$ but not on the potential. For the readers' convenience, we prove that $K_n = \frac{\pi^{\frac{n}2(n-1)}}{n!}$. In order to do so, we explicitly compute the left and the right hand side of \eqref{HCAppendix} in the Gaussian case $V(x)= x^2/2$. 

The left hand side is
\be
\int_{\mathcal H_n} {\rm e}^{-\Tr (M^2/2 - M \Lambda)} \d M =
{\rm e}^{\Tr \Lambda^2/2}\int_{\mathcal H_n} {\rm e}^{-\frac 1 2 \Tr (M-\Lambda)^2} \d M =
{\rm e}^{\Tr \Lambda^2/2} \pi^{n(n-1)/2} (2\pi)^\frac n2
\ee
while the right hand side becomes
\be
 \frac {K_n}{\Delta (\Lambda)} \int_{\R_n} \Delta(X) \det \le[{\rm e}^{x_j \l_k}\ri]_{j,k=1}^{n} \prod_{\ell=1}^n {\rm e}^{-\frac  {x_j^2}2} \d x_j
 \mathop{=}^{\hbox{\tiny Andreief}}
 \frac {n!K_n}{\Delta (\Lambda)} 
 \det \le[
 \int_{\R} {\rm e}^{-\frac {x^2}2 + \l_\ell x} x^{k-1}
 \ri]_{\ell,k=1}^{n}
 =\\
= \frac {n!K_n}{\Delta (\Lambda)} 
 \det \le[
{\rm e}^{\frac {\l_\ell^2}2}  \int_{\R} {\rm e}^{-\frac {(x-\l_\ell)^2}2} x^{k-1}\d x 
 \ri]_{\ell,k=1}^{n}
= \frac {n!K_n {\rm e}^{\frac 1 2 \Tr \Lambda^2} }{\Delta (\Lambda)} 
 \det \le[  \int_{\R} {\rm e}^{-\frac {x^2}2} (x+\l_\ell)^{k-1}\d x 
 \ri]_{\ell,k=1}^{n}
=\\
= \frac {n!K_n {\rm e}^{\frac 1 2 \Tr \Lambda^2} }{\Delta (\Lambda)} 
 \det \le[  \sqrt{2\pi} P_{k-1}(\l_\ell) 
 \ri]_{\ell,k=1}^{n} = 
 {n! K_n (2\pi)^\frac n 2 {\rm e}^{\frac 1 2\Tr \Lambda^2}}
\ee
Here $P_{k-1}(\l)$ is a {\bf monic} polynomial of the indicated degree.
Thus we conclude  that $
K_n  = \frac {\pi^{\frac n 2(n-1)}}{n!}$.

\bibliographystyle{plain}
\def\cprime{$'$}


\begin{thebibliography}{10}



\bibitem{ASvM}
M.~Adler, T.~Shiota, and P.~van Moerbeke.
\newblock Random matrices, {V}irasoro algebras, and noncommutative {KP}.
\newblock {\em Duke Math. J.}, 94(2):379--431, 1998.

\bibitem{AkemannFyodorov}
G.~Akemann and Y.~V. Fyodorov.
\newblock Universal random matrix correlations of ratios of characteristic
  polynomials at the spectral edges.
\newblock {\em Nuclear Phys. B}, 664(3):457--476, 2003.

\bibitem{OxRMBook}
Gernot Akemann, Jinho Baik, and Philippe Di~Francesco, editors.
\newblock {\em The {O}xford handbook of random matrix theory}.
\newblock Oxford University Press, Oxford, 2011.

\bibitem{AndreevSimons}
A.~V. Andreev and B.~D. Simons.
\newblock Correlators of spectral determinants in quantum chaos.
\newblock {\em Physical Review Letters}, 75(12):2304--2307, 1995.

\bibitem{BaikDeiftStrahov03}
J.~Baik P.~Deift and E.~Strahov.
\newblock Products and ratios of characteristic polynomials of random
  {H}ermitian matrices.
\newblock {\em J. Math. Phys.}, 44(8):3657--3670, 2003.
\newblock Integrability, topological solitons and beyond.

\bibitem{BertolaCafassoKonts}
M.~Bertola and M.~Cafasso.
\newblock {The Kontsevich matrix integral: convergence to the Painlev\'e
  hierarchy and Stokes' phenomenon}.
\newblock {\em arXiv: 1603.06420}.

\bibitem{BertolaCafasso5}
M.~Bertola and M.~Cafasso.
\newblock Darboux transformations and random point processes.
\newblock {\em International Mathematics Research Notices}, 15: 6211-6266, 2015.

\bibitem{BrezinHikami00}
E.~Br{{\'e}}zin and S.~Hikami.
\newblock Characteristic polynomials of random matrices.
\newblock {\em Comm. Math. Phys.}, 214(1):111--135, 2000.

\bibitem{BrezinHikami00bis}
E.~Br{{\'e}}zin and S.~Hikami.
\newblock Characteristic polynomials of random matrices at edge singularities.
\newblock {\em Phys. Rev. E (3)}, 62(3, part A):3558--3567, 2000.

\bibitem{BrezinHikamiBessel}
E.~Br{{\'e}}zin and S.~Hikami.
\newblock Duality and replicas for a unitary matrix model.
\newblock {\em J. High Energy Phys.}, (7):067, 16, 2010.

\bibitem{DKMVZ}
P.~Deift, T.~Kriecherbauer, K.~T.-R. McLaughlin, S.~Venakides, and X.~Zhou.
\newblock Uniform asymptotics for polynomials orthogonal with respect to
  varying exponential weights and applications to universality questions in
  random matrix theory.
\newblock {\em Comm. Pure Appl. Math.}, 52(11):1335--1425, 1999.

\bibitem{Desrosiers}
P.~Desrosiers.
\newblock {Duality in random matrix ensembles for all $\beta$}.
\newblock {\em Nuclear Phys. B}, 817(3):224--251, 2009.

\bibitem{DesrosiersLiu}
P.~Desrosiers and D.-Z. Liu.
\newblock Asymptotics for products of characteristic polynomials in classical
  {$\beta$}-ensembles.
\newblock {\em Constr. Approx.}, 39(2):273--322, 2014.

\bibitem{DesrosiersLiubis}
P.~Desrosiers and D.-Z. Liu.
\newblock Scaling limits of correlations of characteristic polynomials for the
  {G}aussian {$\beta$}-ensemble with external source.
\newblock {\em Int. Math. Res. Not. IMRN}, (12):3751--3781, 2015.


\bibitem{FyodorovStrahov03}
Y.~V. Fyodorov and E.~Strahov.
\newblock An exact formula for general spectral correlation function of random
  {H}ermitian matrices.
\newblock {\em J. Phys. A}, 36(12):3203--3213, 2003.
\newblock Random matrix theory.

\bibitem{FIK0}
A.~R. Its, A.~V. Kitaev, and A.~S. Fokas.
\newblock An isomonodromy approach to the theory of two-dimensional quantum
  gravity.
\newblock {\em Uspekhi Mat. Nauk}, 45(6(276)):135--136, 1990.


\bibitem{KeatingSnaith}
J.~P. Keating and N.~C. Snaith.
\newblock Random matrix theory and {$\zeta(1/2+it)$}.
\newblock {\em Comm. Math. Phys.}, 214(1):57--89, 2000.

\bibitem{Kontsevich}
Maxim Kontsevich.
\newblock Intersection theory on the moduli space of curves and the matrix
  {A}iry function.
\newblock {\em Comm. Math. Phys.}, 147(1):1--23, 1992.


\bibitem{MehtaBook}
M.~L. Mehta.
\newblock {\em Random matrices}, volume 142 of {\em Pure and Applied
  Mathematics (Amsterdam)}.
\newblock Elsevier/Academic Press, Amsterdam, third edition, 2004.

\bibitem{KMMM}
A.~Mironov S.~Kharchev, A.~Marshakov and A.~Morozov.
\newblock Generalized {K}azakov-{M}igdal-{K}ontsevich model: group theory
  aspects.
\newblock {\em Internat. J. Modern Phys. A}, 10(14):2015--2051, 1995.

\bibitem{MMS}
A.~Morozov A.~Mironov and G.~W. Semenoff.
\newblock {Unitary matrix integrals in the framework of the generalized
  Kontsevich model}.
\newblock {\em Internat. J. Modern Phys. A}, 11(28), 1996.

\bibitem{DMLF}
{\em NIST Digital Library of Mathematical Functions}.
\newblock http://dlmf.nist.gov/.

\bibitem{OkounkovPandharipandeEdge}
A.~Okounkov and R.~Pandharipande.
\newblock Gromov-{W}itten theory, {H}urwitz numbers, and matrix models.
\newblock In {\em Algebraic geometry---{S}eattle 2005. {P}art 1}, volume~80 of
  {\em Proc. Sympos. Pure Math.}, pages 325--414. Amer. Math. Soc., Providence,
  RI, 2009.

\bibitem{OkounkovKdV}
A. Okounkov.
\newblock Generating functions for intersection numbers on moduli spaces of
  curves.
\newblock {\em Int. Math. Res. Not.}, (18):933--957, 2002.


\bibitem{FyodorovStrahovtris}
E.~Strahov and Y.~V. Fyodorov.
\newblock Universal results for correlations of characteristic polynomials:
  {R}iemann-{H}ilbert approach.
\newblock {\em Comm. Math. Phys.}, 241(2-3):343--382, 2003.

\bibitem{FyodorovStrahov03bis}
E.~Strahov and Y.V. Fyodorov.
\newblock Correlation functions of characteristic polynomials as determinants
  of integrable kernels: universality in the {D}yson limit.
\newblock {\em Markov Process. Related Fields}, 9(4):615--632, 2003.

\bibitem{IshikawaOkada}
H.~Tagawa M.~Ishikawa, S.~Okada and J.~Zeng.
\newblock {Generalizations of Cauchy's determinant and Schur's Pfaffian}.
\newblock {\em Advances in Applied Mathematics}, 36, 2006.

\bibitem{VanlessenUniverBessel}
M.~Vanlessen.
\newblock Universal behavior for averages of characteristic polynomials at the
  origin of the spectrum.
\newblock {\em Comm. Math. Phys.}, 253(3):535--560, 2005.

\end{thebibliography}
\end{document}